\documentclass[aps,10pt,twocolumn,showpacs,preprintnumbers,amsmath,amssymb,showkeys,superscriptaddress]{revtex4-1}

\usepackage{pdftexcmds}
\usepackage[version=3]{mhchem}
\usepackage{siunitx}
\usepackage{xcolor}
\usepackage{graphicx}
\usepackage{dcolumn}
\usepackage{bm}
\usepackage{amsmath}
\usepackage{tikz}
\usepackage{pgfplots}
\usepackage{mathtools}
\usepackage{array}
\usepackage{booktabs}
\usepackage{comment}

\begin{document}
	
\title{Hole conductivity through a defect band in \ce{ZnGa2O4}}
	
\author{Fernando P. Sabino}
\affiliation{Centro de Ci\^encias Naturais e Humanas, Universidade Federal do ABC, 09210-580 Santo Andr\'e, SP, Brazil}

\author{Intuon Chatratin}
\affiliation{Department of Materials Science and Engineering, University of Delaware, 
Newark, Delaware 19716, USA}
	
\author{Anderson Janotti}
\affiliation{Department of Materials Science and Engineering, University of Delaware, 
Newark, Delaware 19716, USA}
	
\author{Gustavo M. Dalpian}
\affiliation{Centro de Ci\^encias Naturais e Humanas, Universidade Federal do ABC, 09210-580 Santo Andr\'e, SP, Brazil}

\begin{abstract}

Semiconductors with wide band gap ($>3.0$~eV), high dielectric constant ($ > 10$), good thermal dissipation, and capable of $n$ and $p$-type doping are highly desirable for high-energy power electronic devices.  Recent studies indicate that \ce{ZnGa2O4} may be suitable for these applications, standing out as an alternative to \ce{Ga2O3}. The simple face centered cubic spinel structure of \ce{ZnGa2O4} results in isotropic electronic and optical properties, in contrast to the large anisotropic properties of the $\beta$-monoclinic \ce{Ga2O3}. 
In addition, \ce{ZnGa2O4} has shown, on average, better thermal dissipation and potential for $n$- and $p$-type conductivity. 
Here we use density functional theory and hybrid functional calculations to investigate the electronic, optical, and point defect properties of \ce{ZnGa2O4}, focusing on the possibility for $p$- and $n$-type conductivity. 
We find that the cation antisite Ga$_{\rm Zn}$ is the lowest energy donor defect that can lead to unintentional $n$-type conductivity. The stability of self-trapped holes (small hole polarons) and the high formation energy of acceptor defects make it difficult to achieve $p$-type conductivity.
However, with excess of Zn, forming \ce{Zn_{(1+2x)}Ga_{2(1-x)}O4} alloys display an intermediate valence band, facilitating $p$-type conductivity.  Due to the localized nature of this intermediate valence band, $p$-type conductivity by polaron hopping is expected, explaining the low mobility and low hole density observed in recent experiments.

\end{abstract}


\maketitle

\section{Introduction} 
\label{sec:Introduction}

Ultra-wide band gap semiconductors (WBS) have attracted great attention in recent years for their application in high-power transistors for energy conversion and solar-blind UV detectors, combining high critical electric field strength, transparency in the visible spectrum, and high conductivity through doping \cite{Tsao_1600501_2018,Gorai_3338_2019,Chen_2079_2020,Chikoidze_2535_2020,Chi_2542_2021}.
Among the promising semiconductor materials, \ce{Ga2O3} stands out with band gap in the range of 4.6-5.0 eV and a large electric breakdown field, resulting in Baliga's figure of merit (BFOM) just smaller than diamond and with capability that goes beyond existent technologies based on \ce{SiC} and \ce{GaN} \cite{Tsao_1600501_2018,Pearton_011301_2018}. 
However, the low thermal conductivity and the complicated monoclinic crystal structure of $\beta$-\ce{Ga2O3} lead to low heat dissipation and anisotropic electronic and optical properties, which, in addition to the inherent difficulties in making the material $p$-type or even reliably semi-insulating, pose serious obstacles to device design and implementation \cite{Pearton_011301_2018,Mastro_P356_2017,Gorai_3338_2019,Chen_2079_2020}.

Recent studies have focused on \ce{ZnGa2O4} as a ultra-wide band gap semiconductor for devices \cite{Gorai_3338_2019,Chen_2079_2020,Chikoidze_2535_2020,Chi_2542_2021}. 
With a band gap of ~5.0 eV, room-temperature electron mobility of $\rm 10^2~cm^2/V.s$, and high dielectric constant of 10.4, \ce{ZnGa2O4} has been considered as an alternative or, at least, as a complement to \ce{Ga2O3} \cite{Chen_2079_2020,Chikoidze_2535_2020,Chi_2542_2021}. 
Thin films of spinel \ce{ZnGa2O4} have been grown epitaxially on sapphire substrate \cite{Chi_2542_2021}. 
Considering an average in all directions, experimental results indicate that the thermal dissipation in \ce{ZnGa2O4} thin films is about 10\% more efficient than in \ce{Ga2O3} \cite{Galazka_022512_2019,Chen_2079_2020}.

Reported unintentional $n$-type conductivity in \ce{ZnGa2O4} with carrier density of $\rm 9x10^{19} cm^{-3}$ is three orders of magnitude higher than that in undoped \ce{Ga2O3} and one order of magnitude higher than in Si- or Sn-doped \ce{Ga2O3} \cite{Ma_212101_2016,Pearton_011301_2018,Chikoidze_2535_2020,Chi_2542_2021}. 
The possible source of this unintentional $n$-type doping in \ce{ZnGa2O4} is still under debate, being attributed to oxygen vacancies ($V_{\rm O}$), antisites such as Ga on Zn site ($\rm Ga_{Zn}$), or a combination of both \cite{Chikoidze_2535_2020,Chi_2542_2021}. 
The oxygen vacancy was predicted to be a shallow donor in some oxides, such as \ce{TiO2} and \ce{In2O3}, for example \cite{Janotti_085212_2010,Chatratin_074604_2019}. 
However, in oxides with wider band gaps, such \ce{Ga2O3} or \ce{ZnO}, the oxygen vacancy was predicted to behave as a deep donor, and could not contribute to the observed unintentional $n$-type conductivity \cite{Janotti_165202_2007,Varley_142106_2010}. 
Similarly, it is expected that the unintentional conductivity observed in \ce{ZnGa2O4} cannot be explained by the presence of $V_{\rm O}$.

Recent experimental results also indicate that \ce{ZnGa2O4} shows $p$-type conductivity when grown in O-rich flux condition \cite{Chikoidze_2535_2020,Chi_2542_2021}, with 
hole concentrations reaching $\rm 10^{15}$~cm$^{-3}$ at 800 K \cite{Chikoidze_2535_2020}, and an extracted activation energy of 0.93 eV.  Such high thermal activation energy was attributed to 
intraband conduction according to E. Chikoidze et. al \cite{Chikoidze_2535_2020}. 
This is an intriguing result considering the very low valence band (high workfunction) of \ce{ZnGa2O4}. 
If both $p$ and $n$-type conductivity could be achieved in the wide band gap \ce{ZnGa2O4}, this material would become a potential candidate for solid-state deep UV LED (light emitting diode), with photon energies in the window of 254-207 nm that is appropriate for eliminating virus and bacterias, including SARS-COV 19 \cite{Buonanno_10285_2020}. 

Although \ce{ZnGa2O4} is already being used as a transparent conducting oxide, the role
of native point defects in their electronic and optical properties is still subject of debate.
Using first principles calculations based on density functional theory (DFT) and hybrid functional, 
calculate the formation energies and transition levels for native point defects in 
\ce{ZnGa2O4}. We find that the $\rm Ga_{Zn}$ antisite is a shallow donor and likely the main source of unintentional $n$-type conductivity in \ce{ZnGa2O4}. On the other hand, none of the native point defects are shallow acceptors, and cannot lead to $p$-type conductivity.  
Moreover, holes tend to become self-trapped, forming small polarons.

On the other hand, if a large excess of Zn is incorporated in \ce{ZnGa2O4}, forming \ce{Zn_{(1+2x)}Ga_{2(1-x)}O4} alloys, the extra Zn occupying the Ga sites lead to a partially occupied intermediate valence band, within which holes existing as small polarons move by hopping, possibly explaining the observed $p$-type conductivity\cite{Chikoidze_2535_2020,Chi_2542_2021}. We also construct configuration coordinate diagrams to determine the energy of the optical emission peaks associated with each defect to help their identification, with special attention to the oxygen vacancy, the self-trapped hole, and the Ga$_{rm Zn}$ antisite.

\section{Theoretical Approach and Computational Details}

Our first principles calculations are based on density functional theory (DFT) \cite{Hohenberg_B864_1964,Kohn_A1133_1965} and the HSE hybrid functional\cite{Heyd_7274_2004,Heyd_219906_2006} as implemented in the Vienna Ab-Initio Simulation Package (VASP) \cite{Kresse_13115_1993,Kresse_11169_1996}.
The interaction between the valence electrons and the ionic cores was treated using  projected augmented wave (PAW) potentials \cite{Blochl_17953_1994,Kresse_1758_1999}, considering the following valence configurations: \ce{O}($2s^{2} 2p^{4}$), \ce{Ga}($3d^{10} 4s^{2} 4p^{1}$) and \ce{Zn}($4d^{10} 5s^{2}$).
In the HSE functional, the exchange is separated in two regions, long and short range, by a screening parameter $\omega = 0.20$ \cite{Heyd_7274_2004,Heyd_219906_2006}.
In the default configuration, the short range region is composed of 25\% of non-local Hartree-Fock exchange, and the remaining 75\% derives from the Perdew-Burke-Ernzerhof (PBE) functional. All the correlation and the exchange for the long-range region are derived from the PBE functional \cite{Heyd_7274_2004,Heyd_219906_2006}. However, the default value for the amount of Hartree-Fock exchange is not enough to correct the band gap for both \ce{Ga2O3} and \ce{ZnO}, and we expect the same behavior for \ce{ZnGa2O4}. Here we find that a mixing parameter of 33\% is necessary to give a band gap of 4.9 eV for \ce{ZnGa2O4}, which leads to a good agreement to recent experimental data \cite{Chikoidze_2535_2020,Chi_2542_2021}. 

The stress tensor and the atomic forces were minimized for the primitive spinel structure of \ce{ZnGa2O4} within the HSE functional using a plane wave cutoff energy of 600 eV. For the Brillouin zone integration, we employed a \textbf{k}-mesh of 5$\times$5$\times$5, and for the density of states (DOS) we increased the mesh to 9$\times$9$\times$9. The formation energies of defects and the emission spectrum were computed using a supercell that is a 2$\times$2$\times$2 repetition of the primitive spinel, totaling a structure with 112 atoms.

\subsection*{Defect formation energies}

\begin{figure}
\centering
\includegraphics[width=1.00\linewidth]{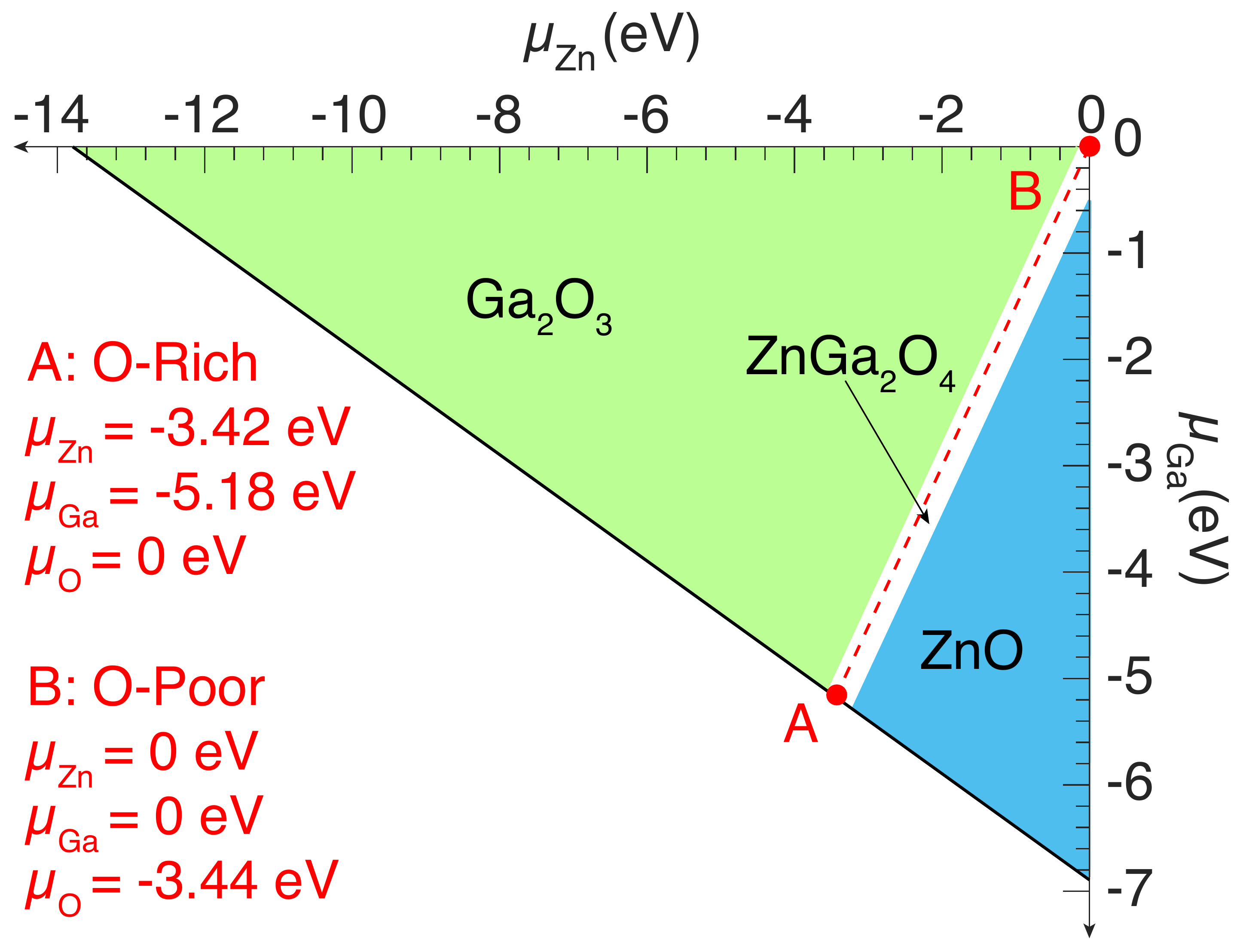}
\caption{
Phase diagram (or stability triangle) for \ce{ZnGa2O4}. The white area represents the chemical potentials that lead \ce{ZnGa2O4} with higher stability when compared to the unary oxides of \ce{ZnO} and \ce{Ga2O3}. Points A and B represent the condition of O-rich and O-poor, respectively, and the chemical potentials for O, Zn, and Ga are shown for both points. 
}
\label{phase}
\end{figure}

\begin{figure*}
\centering
\includegraphics[width=0.80\linewidth]{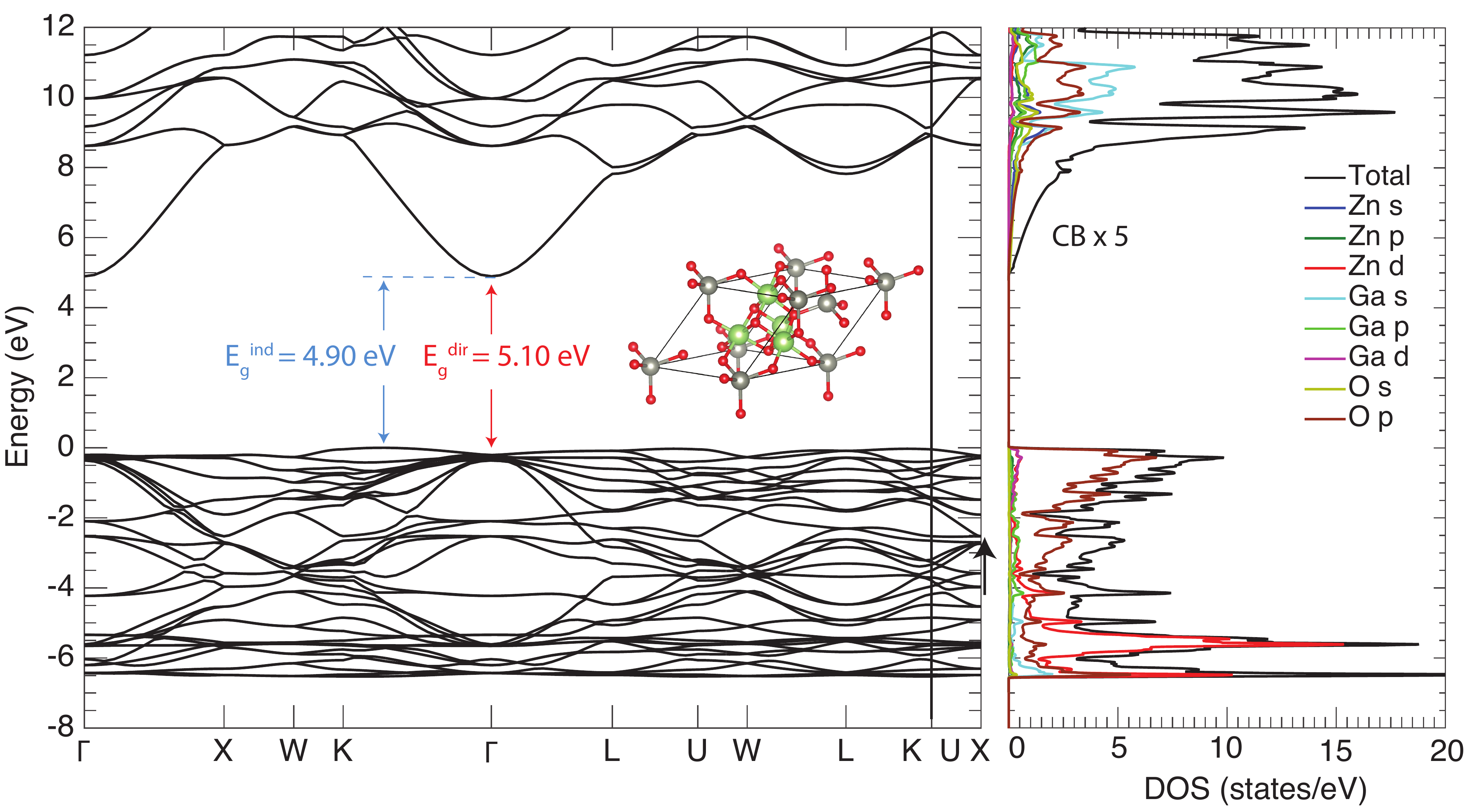}
\caption{
Band structure and density of states for the FCC spinel structure of \ce{ZnGa2O4} taking the VBM as a reference. The band gap is indirect with the VBM located along the  $\Gamma$-K direction, and the CBM occurs at $\Gamma$. The crystal structure of \ce{ZnGa2O4} where the Zn, Ga, and O are shown by the gray, green, and red spheres, respectively, is represented in the inset. 
}
\label{band}
\end{figure*}

The formation energy of a defect $D$ in the charge state $q$ was calculated according to the following expression:  
\begin{equation}
 \label{eq:EA}
 E^f(D,q) = E_{tot}(D,q) - E_{tot}(bulk) + \sum_i n_i\mu_i+qE_F+\Delta^q,
\end{equation}
where $E_{tot}(D,q)$ is the total energy of the supercell containing a defect $D$ in charge state $q$, and $E_{tot}(bulk)$ is the total energy of perfect \ce{ZnGa2O4} in the same supercell.  
The chemical potential $\mu_i$ for the specie $i$ that was removed or added to the supercell for the defect is referenced to the total energy of the respective elemental phase.  
$E_F$ is the Fermi energy taking the valence-band maximum (VBM) as a reference, and $\Delta^q$ is the charge-state dependent correction due to the finite size of the supercell \cite{Freysoldt_016402_2009,Freysoldt_1067_2011}. 

The chemical potentials are treated as variables in Eq.~(1). To ensure the stability of \ce{ZnGa2O4}, the chemical potentials should satisfy the following condition:  
\begin{equation}
 \label{eq:EA}
 \mu_{Zn} + 2\mu_{Ga} + 4\mu_{O} = \Delta H_f({\rm ZnGa_2O_4}), 
\end{equation}
where $\Delta H_f({\rm ZnGa_2O_4})$ is the formation enthalpy of \ce{ZnGa2O4}; we find  $\Delta H_f({\rm ZnGa_2O_4}) = -13.79$~eV using HSE. To avoid the formation of competing phases, such as \ce{Ga2O3} and \ce{ZnO}, the chemical potentials must also satisfy:
\begin{equation}
\begin{multlined}
 \label{eq:EA}
 2\mu_{Ga} + 3\mu_{O} < \Delta H_f({\rm Ga_2O_3)}, \\
 \mu_{Zn} + \mu_{O} < \Delta H_f({\rm ZnO}),
 \end{multlined}
\end{equation}
where $\Delta H_f({\rm Ga_2O_3)}$ and $\Delta H_f({\rm ZnO})$ are the formation enthalpies of $\beta$-\ce{Ga2O3} and wurtzite \ce{ZnO}. Following these relations, we can determine the region of stability for \ce{ZnGa2O4} in the phase diagram as shown in Fig.~\ref{phase}. The blue and green areas show the regions where the oxides \ce{ZnO} and \ce{Ga2O3} are stable, while the 
white area represents the region where \ce{ZnGa2O4} is stable. In the phase diagram, we choose two points, A and B, for which the defect formation energies are reported, representing O-rich and O-poor limit conditions, respectively.

The optical emission spectrum for point defects was determined based on calculated configuration coordinate diagrams, where it is assumed that the optical excitations occur at fixed atomic coordinates, as reported for \ce{In2O3} and \ce{Ga2O3} in the literature \cite{Chatratin_074604_2019,Zimmermann_072101_2020}.

\section{Results and discussion}

\subsection{Crystal structure and electronic properties}

\ce{ZnGa2O4} crystallizes in the face center cubic (FCC) spinel structure that belongs to $Fm\overline{3}m$ space group with an experimental equilibrium lattice parameter of $a_0 = 8.33$~$\rm \AA$ \cite{Chikoidze_2535_2020} (the crystal structure is shown in the inset of Fig.~\ref{band}). 
The calculated equilibrium lattice parameter of 8.36~\AA~ is only  $0.36\%$ larger than the experimental value. In the spinel crystal structure, the oxygen atoms occupy a unique non-equivalent site with Wyckoff notation 32e, where each O is bound to four cations (three Ga and one Zn). 
On the other hand, there are two non-equivalent cation sites, with completely different chemical environments.
One of these sites is surrounded by six oxygen atoms, forming perfect octahedral motifs (16d in the Wyckoff notation), while the second site forms a perfect tetrahedral chemical environment (8a in the Wyckoff notation). 
All the local chemical environments are shown in Fig.~S1 in the Supplemental Material.

The electronic band structure and density of states of \ce{ZnGa2O4}, calculated using HSE, are shown in Fig.~\ref{band}. 
The valence band is mainly derived from O 2$p$-orbitals, with the VBM located in between the K and $\Gamma$ points; it shows low dispersion (high effective masses) and is low in energy (high ionization potential). 
In contrast, the conduction band is composed mainly of Zn and Ga $s$-orbitals, resulting in a large dispersion (CBM), with a minimum at $\Gamma$. The calculated indirect band gap is 4.90 eV. The local maximum in the valence band at $\Gamma$ lies only 0.20 eV below the VBM, leading to a direct band gap of 5.10 eV. 

The dispersion of the VBM in \ce{ZnGa2O4} is affected by the $p$-$d$ coupling. 
In \ce{ZnO}, the coupling between Zn $d$ and O $p$ orbitals shifts the position of the VBM to higher energies, increasing the dispersion in the vicinity of VBM \cite{Janotti_165202_2007}. 
The position of Ga $d$-orbitals in \ce{Ga2O3}, on the other hand, is significantly lower than the VBM, resulting in a weaker $p$-$d$ coupling and a low dispersion of the VBM \cite{Sabino_155206_2014,Sabino_205308_2015}.  In \ce{ZnGa2O4}, the $p$-$d$ coupling has an intermediate intensity, lying between the cases of \ce{ZnO} and \ce{Ga2O3}, with contributions from both cations $d$-orbitals. 
Zn $d$ orbitals are located around 4 eV below the VBM, as indicated by the red line in the orbital-resolved DOS in Fig.~\ref{band}, while Ga 3$d$-orbitals lies 14 eV below the VBM (not shown in Fig.~\ref{band}). 
Despite the moderated $p$-$d$ coupling, the intensity is not sufficient to increase the dispersion of the VBM in \ce{ZnGa2O4}.

\subsection{Intrinsic defects}

We investigate all the possible intrinsic point defects in \ce{ZnGa2O4}, i.e., vacancies ($\rm V_O$, $\rm V_{Zn}$ and $\rm V_{Ga}$), 
antisites ($\rm Ga_{Zn}$, $\rm Ga_{O}$, $\rm Zn_{Ga}$, $\rm Zn_{O}$, $\rm O_{Zn}$ and $\rm O_{Ga}$), and interstitials ($\rm Zn_{i}$, $\rm Ga_{i}$ and $\rm O_{i}$), as well as the self-trapped hole in the form of a small hole polaron. The defect formation energies as a function of Fermi level for O-rich (point A) and O-poor (point B) conditions are shown in Fig.~\ref{phase}.

For Fermi level near the valence band and for the O-rich condition, the most stable defects are $\rm Ga_{Zn}$, $\rm Ga_{i}$, $\rm Zn_{i}$ and $\rm V_{O}$; they are all donor defects. In contrast, for Fermi level near the conduction band, $\rm V_{Zn}$, $\rm V_{Ga}$ and $\rm Zn_{Ga}$ are the lowest energy defects and act as acceptors. For the O-poor condition, the results are similar, however, the donors $\rm Ga_{O}$ and $\rm Zn_{O}$ are also low formation energy defects for Fermi level below the middle of the band gap. 

It is important to mention that, in O-rich conditions, the acceptor defects are predominant in the region close to the CBM. 
The formation energy of these acceptor defects are significantly lower than any donor defect and, therefore, we expect $n$-type conductivity under O-rich condition to be very difficult to be achieved due to compensation. On the other hand, acceptor defects are high formation energy defects in the region near the VBM for O-poor condition, preventing, in principle, $p$-type conductivity. We also note that the lowest energy defects that can cause $n$-type ($p$-type) conductivity is $\rm Ga_{Zn}$ ($\rm Zn_{Ga}$), i.e. the cation antisites. This explains the observation of Ga (Zn) occupying the tetrahedral (octahedral) site in the spinel structure \cite{Chen_2079_2020,Chikoidze_2535_2020,Chi_2542_2021}.
The most likely source of unintentional $n$-type conductivity in \ce{ZnGa2O4} is $\rm Ga_{Zn}$, whereas $\rm V_{O}$, often pointed as the source of unintentional $n$-type conductivity in many oxides, is a deep donor.  In the following, we discuss the details of each defect in \ce{ZnGa2O4}, separating the discussion on hole polaron, vacancies, antisites, and interstitial defects. 

\begin{figure}
\centering
\includegraphics[width=1.00\linewidth]{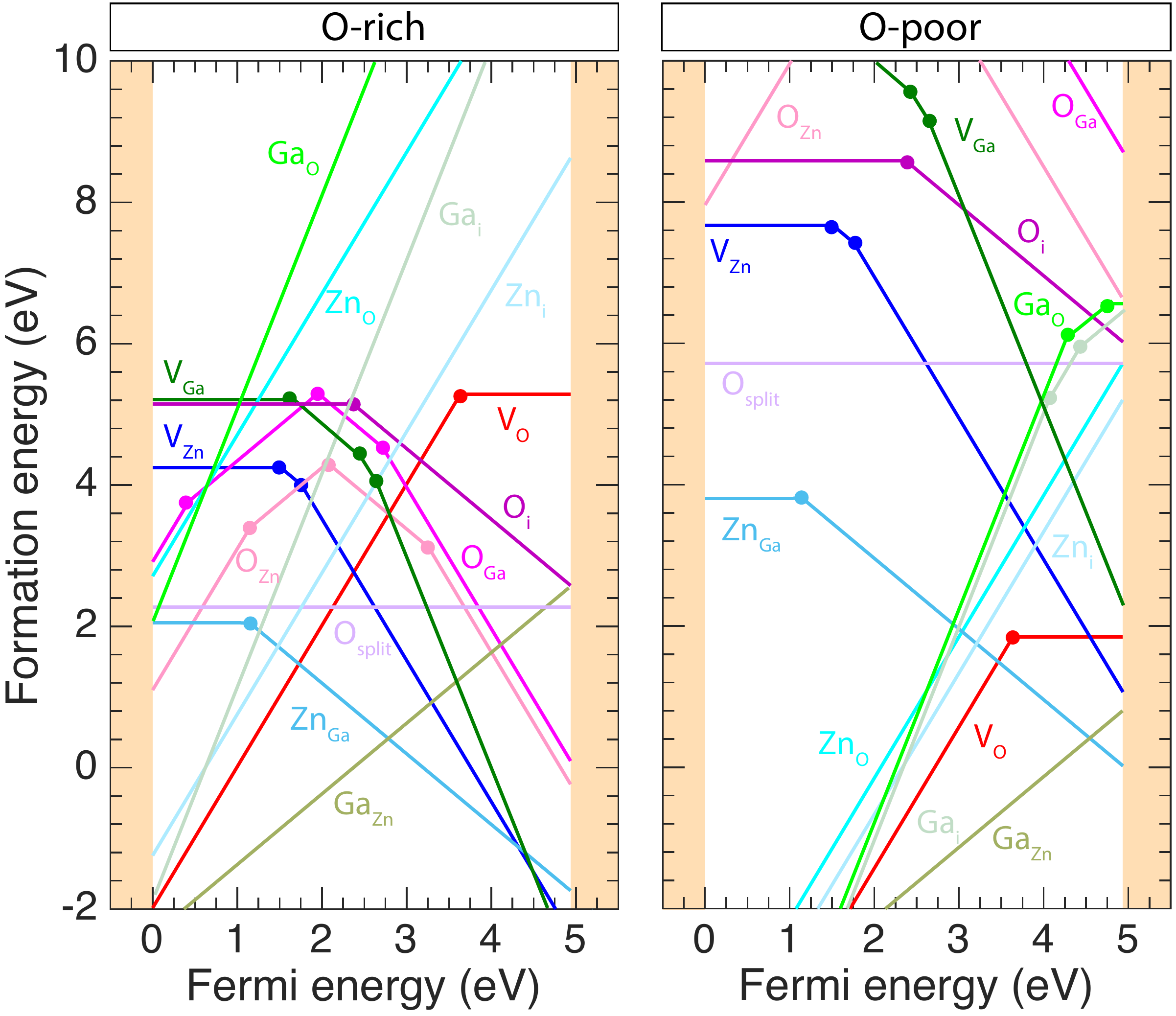}
\caption{
Formation energy as a function of the parametric Fermi energy for all intrinsic defects in \ce{ZnGa2O4} 
calculated in O-rich and poor chemical potentials, which is represented in Fig.~\ref{phase}. The defects 
that are associated with oxygen are represented in red tons, the ones associated with zinc in blue, and the ones for Ga in green.  
}
\label{formation}
\end{figure}

\subsubsection{Small hole polaron}

We find that a hole in the valence band of \ce{ZnGa2O4} spontaneously becomes self-trapped, forming a small hole polaron, as predicted for other oxides \cite{Valey_081109_2012,Gake_044603_2019,Lyons_025701_2022}.  It is localized on one oxygen atom and accompanied by a local lattice distortion where one O-Ga bond length increased by 23\%, transforming the tetrahedral environment around O to an almost trigonal planar configuration, with the wavefunction resembling that of an O 2$p_z$ orbital as shown in Fig.~\ref{structure}~(a). This effect is caused by the rather low dispersion and low energy of the valence band in \ce{ZnGa2O4}. The calculated hole self-trapping energy of 0.44 eV is essentially defined as the difference between the total energy of the supercell containing one localized hole (by removing an electron from the supercell and letting the atomic positions to relax) and the supercell with a hole in the VBM. The relatively large self-trapping energy (much higher than $k_BT$ at room temperature) indicates that hole conductivity, if realized, will be given by small polaron hopping, with characteristic low mobility at room temperature.

\subsubsection{Vacancies}

\begin{figure}
\centering
\includegraphics[width=1.00\linewidth]{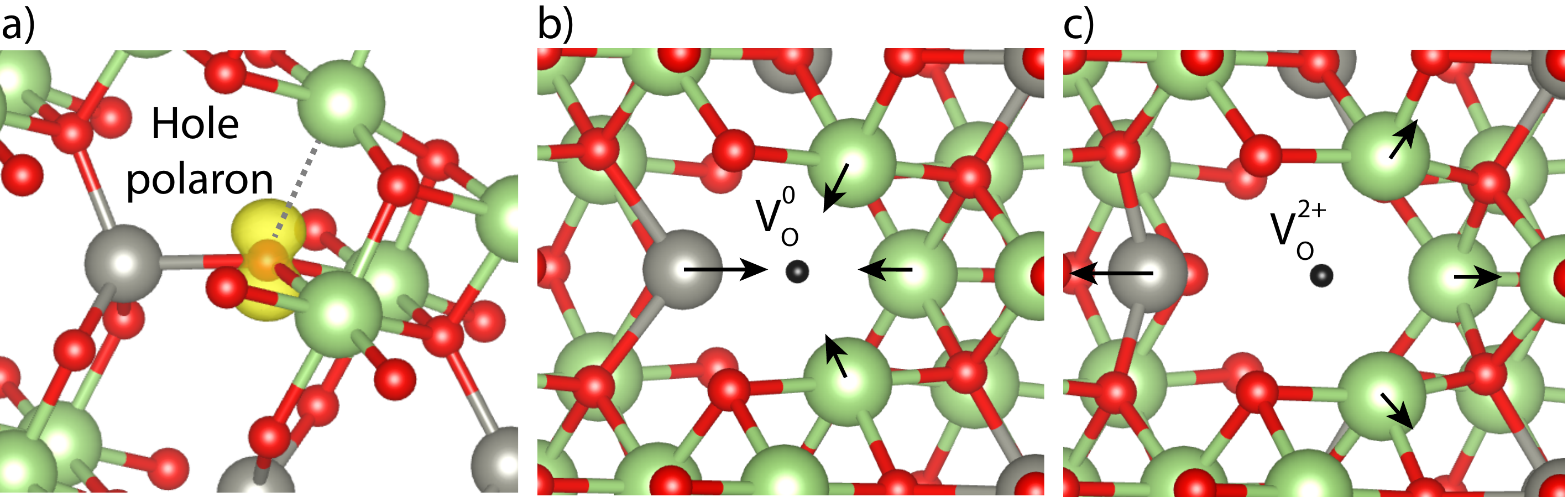}
\caption{
Local atomic environment representing the a) hole polaron, with the local charge density (yellow); 
b) oxygen vacancy in the neutral charge state, $\rm V_O^0$; and
c) oxygen vacancy in the 2+ charge state, $\rm V_O^2+$. The arrows indicate the local lattice relaxations with respect to 
the system without defects. The oxygen, zinc, and gallium atoms are represented by the red, gray, and green colors, respectively.
}
\label{structure}
\end{figure}

$\it Oxgen~vacancy$. When an oxygen vacancy is created in \ce{ZnGa2O4}, four bonds are broken (three with Ga and one with Zn) leaving four dangling bonds. This electronic configuration reorganizes in a low energy single-particle state, and a higher energy threefold degenerate state. In the neutral charge state, the lower one fold degenerate state is occupied with two electrons and located in the band gap, 2.47 eV above the VBM, while the threefold state is unoccupied and lies resonant in the conduction band. Under this charge state, the atoms in the vicinity of $\rm V_O$ relax inward by 14.51\% of the equilibrium Zn-O bong length and the Ga atoms relax inward by 0.50\% of the equilibrium Ga-O bond length, as indicated in Fig.~\ref{structure}(b). 
Removing one electron from the gap state leads to 1+ charge state, $\rm V_O^{1+}$; the gap state splits into spin-up occupied and spin-down unoccupied states, located at 2.90 eV and 4.53 eV above the VBM, respectively. The neighboring atoms relax outward from the vacancy by 6.21\% and Ga by 5.90\% of the respective equilibrium bond length. In the 2+ charge state, the lower state is also completely unoccupied and is shifted up to inside the conduction band. Under this charge state, the neighboring atoms further relax outward: the Zn by 30.05\% and the three Ga 11.38\%, as shown in Fig.~\ref{structure}(c). The neighboring Zn atom around the $\rm V_O^{2+}$ forms an almost trigonal planar motif.

The formation energies as a function of the Fermi level ($E_F$) for $\rm V_O$ in different charge states are shown in Fig.~\ref{formation}. We observe that $\rm V_O^{1+}$ is not stable for all possible values of $E_F$ when compared to the neutral and 2+ charge states. This indicates that the gap state is either fully occupied with two electrons or empty resonant in the conduction band. The thermodynamic transition level from neutral to 2+ charge state occurs at 1.26 eV below the CBM, making $\rm V_O$ a deep donor and unlikely to contribute to $n$-type conductivity at room temperature.  On the other hand, if the equilibrium Fermi level lies in the middle of the band gap (insulator) or close to the VBM ($p$-type), $\rm V_O$ acts as a compensation defect for holes. The low or negative formation energy of $\rm V_O$ in O-poor condition, second lower between all the intrinsic donor defects, make it very difficult to make \ce{ZnGa2O4} $p$-type.  

Therefore, in addition to the high stability of hole polaron discussed before, this is another constable to realize $p$-type conductivity in \ce{ZnGa2O4}. 
In contrast, $\rm V_O$ also does not act as a source of free electrons when the Fermi level in \ce{ZnGa2O4} is close to the CBM and the $n$-type conductivity is predominant.
This behavior is in contradiction with that in \ce{In2O3} or \ce{TiO2} \cite{Janotti_085212_2010,Chatratin_074604_2019}, but similar to that in  ZnO\cite{Janotti_165202_2007} and \ce{Ga2O3}\cite{Varley_142106_2010}.

$\it Zinc~vacancy$. The removal of a Zn atom results in two holes in the valence band that tend to localize in the form of small hole polarons at the O 2$p_z$ orbital of two different O atoms neighboring the vacancy ($\rm V_{Zn}^0$). Because of the electrostatic repulsion of positive charges, the distance between these two hole polarons is maximized around the Zn vacancy; these small polarons are associated with unoccupied single-particle states in the gap, in opposite spin channels.

By adding one electron to $\rm V_{Zn}^0$, one of the small hole polarons vanishes, resulting in the negatively charged state $\rm V_{Zn}^-$, with a (0/-) transition level at 1.51~eV above the VBM. Introducing a second electron fills the remaining single-particle level of $\rm V_{Zn}^{1-}$ inside the band gap, resulting in $\rm V_{Zn}^{2-}$. The ($-$/2$-$) transition level is at 1.76 eV above the VBM. 
The local lattice configuration around $\rm V_{Zn}^{2-}$ is shown in Fig.~S2 in the Supplemental Material.  

The formation energy of $\rm V_{Zn}$ is low, and even becomes negative in O-rich conditions when $E_F$ is near the CBM. This indicates that $\rm V_{Zn}$ is a likely source of electron compensation in $n$-type \ce{ZnGa2O4}. For $E_F$ near the VBM, $\rm V_{Zn}$ has high formation energy. The position of the (0/$-$) and ($-$/2$-$) transition levels well above the CBM indicates that $\rm V_{Zn}$ is a deep acceptor and cannot serve as a source of $p$-type conductivity.  

$\it Gallium~vacancy$. The vacancy of Ga is similar to the Zn. Because of the 3+ oxidation state of Ga, three holes are generated in the system when a $\rm V_{Ga}$ is created. These three holes tend to localize in the form of small polarons and sit on three different O 2$p_z$ orbitals around the $\rm V_{Ga}$, maximizing the distance between them. Because the large electrostatic repulsion between these polarons, $\rm V_{Ga}^0$, is less stable than $\rm V_{Zn}^{0}$. When an electron is added to the system, one semi-occupied quasi-particle state that lies in the band gap (neutral charge state) become completely occupied and move toward the valence band. This condition represents the $\rm V_{Ga}^{1-}$ charge state and the transition from (0/1-) occur at 1.63~eV above the VBM. Similar behavior is observed for the other negative charge states, with transitions (1-/2-) and (2-/3-) lying at 2.40~eV and 2.53~eV above the VBM. All these transitions can be observed in the formation energy of Fig.~\ref{formation} and the atomic configuration for $\rm V_{Ga}^{3-}$ is shown in Fig.~S2 in the Supplemental Material.

In O-rich limit condition, the formation energy of $\rm V_{Ga}$ becomes low or negative for $E_F$ in the upper part of the gap, while in O-poor condition this defect has very high formation energy. In the O-rich limit, the Ga vacancy is stable in the 3$-$ charge state, $\rm V_{Ga}^{3-}$, acting as compensation center to $n$-type conductivity.

\subsubsection{Antisites}

As mentioned before, the antisites, especially $\rm Zn_{Ga}$ and $\rm Ga_{Zn}$, play an important role in the electronic conductivity of \ce{ZnGa2O4}.
Both chemical environments are shown in .
While $\rm Ga_{Zn}$ is the lowest energy donor defect and a likely source of unintentional $n$-type conductivity, $\rm Zn_{Ga}$ is the lowest energy acceptor defect that acts as an electron compensation center. 

We can define the Fermi level pinning for $n$- and $p$-type conductivity, $E_{pin}^{n}$ and $E_{pin}^{p}$, respectively, as the maximum and minimum position for $E_F$ in the gap, taking the VBM as a reference, before the intrinsic defect starts to compensate the prevalent conductivity in the material \cite{Zhang_1232_2000,Zunger_57_2003,Wei_337_2004,Sabino_2022}. 
For materials that are are easy to dope $n$-type, $E_{pin}^{n}$ lies near the edge or resonant in the conduction band. Conversely, for materials that can easily doped $p$-type, $E_{pin}^{p}$
lies in the near or below the VBM. If the Fermi level pinning for $n$- and $p$-type doping lie near the middle of a wide band gap, the material tend to show insulating behavior.  From Fig.~\ref{formation}, we can conclude that in O-rich limit, \ce{ZnGa2O4} will tend to show insulating behavior, with Fermi level pinning near the middle of the gap, while in O-poor limit, it will tend to show $n$-type conductivity, with Fermi level pining near the CBM, in both cases dictated by $\rm Ga_{Zn}$ and $\rm Zn_{Ga}$. 

In the following we discuss each antisite defect in detail; the local lattice relaxations associated with $\rm Zn_{Ga}$ and $\rm Ga_{Zn}$ are shown in Fig.~\ref{antisite} and the remain antsites, with less importance for this material, are shown in Fig.~S2 in the Supplemental Material.

\begin{figure}
\centering
\includegraphics[width=1.00\linewidth]{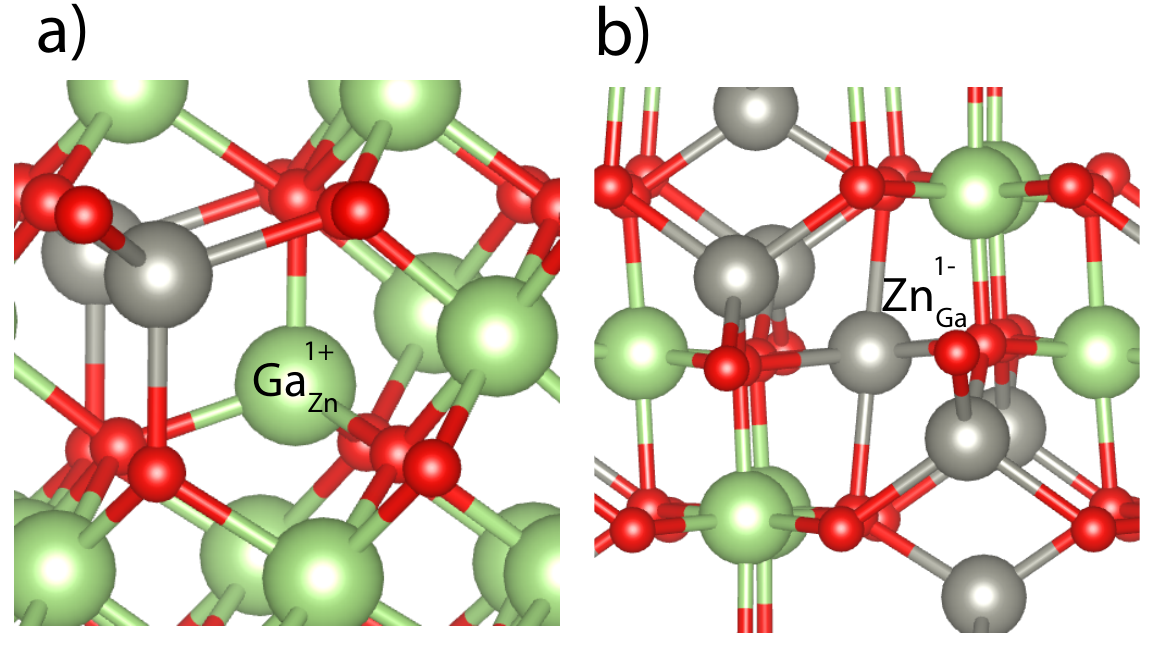}
\caption{
Local chemical environment for antisites defects a) Gallium substitutional on Zinc site in 1+ charge state and, b) Zinc substitutional on gallium site in 1- charge state. The oxygen, zinc, and gallium atoms are represented by the red, gray, and green colors, respectively.
}
\label{antisite}
\end{figure}

$\it Gallium~on~zinc~site$. In the case of $\rm Ga_{Zn}$, Ga with 3+ formal oxidation state replaces Zn with 2+ formal oxidation state in the tetrahedral sites (8a in the Wyckoff notation). This aliovalent substitution creates one unpaired electron that is weakly bound to Ga. The neutral charge $\rm Ga_{Zn}^0$, therefore, unstable, making this defect a shallow donor, with $\rm Ga_{Zn}^{1+}$ being the lowest energy charge state for all values of Fermi level in the gap. In the tetrahedral chemical environment of $\rm Ga_{Zn}^{1+}$, shown in Fig~\ref{antisite}a), $\rm Ga_{Zn}$-O bond length is 1.88~$\rm \AA$, i.e., 2.59\% smaller than the original Zn-O bond length of 1.93~$\rm \AA$.

Among all the intrinsic donor defects, the formation energy of $\rm Ga_{Zn}^{1+}$ is the lowest in both O-rich and O-poor conditions. These results have two consequences: (i) if \ce{ZnGa2O4} exhibits unintentional $n$-type conductivity or is semi-insulator, the main source of electrons in the conduction band or hole compensation is $\rm Ga_{Zn}$; (ii) the low formation energy of this defect indicates that Ga can incorporate in the tetrahedral sites in large concentrations, as observed in recent experiments \cite{Chen_2079_2020,Chikoidze_2535_2020,Chi_2542_2021}. 

The formation energy of $\rm Ga_{Zn}$ also determines $E_{pin}^{p}$, which is given by the point where the formation energy of $\rm Ga_{Zn}^{1+}$ is equal to zero in the O-rich condition with the richest Zn chemical potential, which is not exactly the A point used to plot the Fig.~\ref{formation}. If the Fermi level moves from $E_{pin}^{p}$ towards the VBM, $\rm Ga_{Zn}$ will have negative formation energy, and it will form spontaneously. Under this condition, holes will be automatically compensated by the electrons from $\rm Ga_{Zn}$. Therefore, Fermi level pinning at $\sim$2~eV above the VBM and the high stability of the small hole polaron makes it very difficult to realize $p$-type conductivity in \ce{ZnGa2O4}.

$\it Zinc~on~gallium~site$. In this case, Zn substitutes on the octahedral Ga site (16d in Wyckoff notation). In the neutral charge state, $\rm Zn_{Ga}$ leaves an unpaired electron in the valence band that becomes a localized hole in the form of a small polaron sitting on a neighboring O atom. More precisely, the hole is derived from a hybridization of Zn $d_{x^2-y^2}$ with O $p$ orbitals, as shown in Fig.~S3 in the Supplemental Material, and differ from the polaron previously discussed and shown in Fig.~\ref{structure} (a). 
The hole localization, in this case, is accompanied by a local lattice relaxation in which the $\rm Zn_{Ga}$-O bond length is shorter 4.3\% than the original Ga-O bond length. By adding one electron to form $\rm Zn_{Ga}^{1-}$, the hole polaron vanishes, and the octahedral environment $\rm Zn_{Ga}$, shown in Fig.~\ref{antisite}b), becomes more symmetric. The $\rm Zn_{Ga}$-O bond length of 2.10 \AA~ is 3.96\% longer than the original Ga-O bond length of 2.02 \AA.

As shown in Fig.~\ref{formation}, neutral $\rm Zn_{Ga}^{0}$ is stable only for $E_F$ near VBM. However, its formation energy is high in both O-rich and O-poor conditions. The negative charge state, $\rm Zn_{Ga}^{1-}$, is stable for $E_F$ larger than 1.15~eV, marking the (0/$-$) thermodynamic transition level. $\rm Zn_{Ga}^{1-}$ has the lowest formation energy among all the acceptor intrinsic defects in the O-poor condition and is comparable to $\rm V_{Zn}^{2-}$ and $\rm V_{Ga}^{3-}$ in the O-rich condition for $E_F$ near the CBM. Therefore, we predict that $\rm Zn_{Ga}$ is one of the main sources of electron compensation in \ce{ZnGa2O4}, determining $E_{pin}^{n}$. The pinning Fermi level for the electrons is calculated at the point where $\rm Zn_{Ga}^{1-}$ have its formation energy equal to zero in the O-poor with Ga richest chemical potential condition (which is not exactly the point B used to plot the Fig.~\ref{formation}), and it lies 0.23 eV above the CBM (inside the conduction band). In other words, for $E_F > E_{pin}^{n}$, electrons added to the material will fill the holes associated with $\rm Zn_{Ga}$. Having $E_{pin}^{n}$ near CBM indicates that \ce{ZnGa2O4} can be easily doped $n$-type, and that if doped with impurities, high electron concentrations can eventually be attained, even compared to \ce{Ga2O3}, as reported by recent experiments \cite{Chikoidze_2535_2020}.

$\it Cations~on~oxygen~site$. The substitution of O for Zn or Ga ($\rm Zn_{O}$, $\rm Ga_{O}$) in the tetrahedral environment (32e in Wyckoff notation), create an excess of electrons and a large Coulomb repulsion between the atoms around it. This repulsion is so strong that repels one Zn atom from the original lattice in the tetrahedral (8a) to the interstitial site (16c); the neighboring Zn atom relaxes from a tetrahedral motif to an octahedral environment as shown in Fig.~S2 in the Supplemental Material. 
The displacement of the \ce{Zn} atom is needed to accommodate the much larger $\rm Zn_{O}$ or $\rm Ga_{O}$ atoms, compared to the size of the original O atom.

In the formation energy plot shown in Fig.~\ref{formation}, Zn in O site is stable in the 2+ charge $\rm Zn_{O}^{2+}$. On the other hand, $\rm Ga_{O}$ is stable in three charge states, 3+, 1+ and 0, with transitions (3+/1+) 0.64~eV and (1+/0) 0.24~eV below the CBM. If we consider $E_F$ near the CBM ($n$-type conductivity), all of these defects have very high formation energy, regardless of the O chemical potential. This indicates that $\rm Zn_{O}$and $\rm Ga_{O}$ cannot be a possible source of electrons in $n$-type \ce{ZnGa2O4}. The formation energy of $\rm Zn_{O}$ and $\rm Ga_{O}$ is low or negative only for $E_F$ below the middle of the band gap. However, as noted before, $E_{pin}^{p}$ does not allow $E_F$ to go to this region, and therefore, in general, we can conclude that the concentration of cations in the O site would be negligible in \ce{ZnGa2O4}. 

$\it Oxygen~on~cation~site$. We tested the possibility of O substituting on Zn and Ga sites, namely $\rm O_{Zn}$ and $\rm O_{Ga}$, in \ce{ZnGa2O4}. In the relaxed configuration, $\rm O_{Ga}$ binds with neighboring O atom, forming a $\rm O_{Ga}$-O split interstitial, with a bond length of 1.34~$\rm \AA$ as shown in Fig.~S2 in the Supplemental Material. For $\rm O_{Zn}$, the $\rm O_{Zn}$-O bond length is 1.41~$\rm \AA$. However, in this configuration, both O atoms bond to two Ga neighbors, likely due to the small internal volume of the 8a tetrahedral compared to the 16d octahedral. 

As seen in Fig.~\ref{formation}, $\rm O_{Zn}$ and $\rm O_{Ga}$ can act as donors or acceptors. If $E_F$ lies close to the VBM (CBM) the most stable charge defect is the 2+ (2-), being amphoteric. Note that the neutral charge state is not stable for both defects, and a negative-U behavior is observed with a  transition from 1$-$ to 1+ at 2.07~eV and 1.95~eV above the VBM for $\rm O_{Zn}$ or $\rm O_{Ga}$, respectively. Similar to the cation on the O site, 
the formation energies of $\rm O_{Zn}$ and $\rm O_{Ga}$ are very high in the possible range of $E_F$ determined by $E_{pin}^{n}$ and $E_{pin}^{p}$. As consequence, we do not expect large concentrations of $\rm O_{Zn}$ and $\rm O_{Ga}$ in \ce{ZnGa2O4}. 

\subsubsection{Interstitials}

There are two non-equivalent interstitial sites in the spinel crystal structure, labeled 16c and 8b according to the Wyckoff notation. Both
are shown in Fig.~S1 in the Supplemental Material. The 16c interstitial site is surrounded by six equidistant O atoms at 2.16~$\rm \AA$, forming an octahedral motif. The 8b site is surrounded by four equidistant O at 1.67~$\rm \AA$ and four Ga at 1.81~$\rm \AA$. We tested both 16c and 8b sites for Zn, Ga, and O, the formation energies are shown in Fig.~\ref{formation}. Because of the small volume of the 8b site, it is unlikely that Zn or Ga will occupy this interstitial site, leading to formation energies that are significantly higher than Zn Or Ga occupying 16c site. The local lattice relaxation associated with the interstitial defects, for the most stable charge states, are shown in Fig.~S2 in the Supplemental Material.

$\it Cations~interstitial$. As noted, the most stable interstitial site for $\rm Ga_{i}$ and $\rm Zn_{i}$ is the 16c. In both cases, the interstitial cation is arranged in an octahedral environment with Ga-O bond length that ranges from 1.94~$\rm \AA$ to 2.04~$\rm \AA$ while for Zn-O, from 2.10~$\rm \AA$ to 2.11~$\rm \AA$. In addition, $\rm Ga_{i}$ and $\rm Zn_{i}$ disturb the local chemical environment of the first Zn neighbors, moving them outward from the interstitial. This repulsion is stronger for $\rm Ga_{i}$ and is observed in the local atomic environment shown in Fig.~S2 in the Supplemental Material. The ionized defect composes the most stable configuration. In the case of Zn (Ga), 2 electrons of $s$-orbital (2 from $s$ and 1 from $p$) are weekly bound, so that these interstitials are most stable in the 2+ and 3+ charge states, acting as donors.

In Fig.~\ref{formation} for both O-rich and O-poor conditions, $\rm Ga_{i}$ and $\rm Zn_{i}$ have very high formation energies when $E_F$ is near the CBM; therefore, we do not expect that these defects contribute to the observed unintentional $n$-type conductivity. Only in the region with $E_F$ below the middle of the band gap, $\rm Ga_{i}$ and $\rm Zn_{i}$ have low formation energies, being a potential source of hole compensation in \ce{ZnGa2O4}. This behavior is very similar to the cation on the O site, and especially in O-poor conditions, where their formation energies are very similar. 
 
$\it Oxygen~interstitial$. We tested the O interstitial in both 16c and 8b Wyckoff positions. For $\rm O_{i}$ in the 8b position, we observed a special condition that we call $\rm O_{splitt}$, i.e., the interstitial O binds to a native O, forming an O-O with a bond length of 1.41~$\rm \AA$, suggesting a formation of an \ce{O2} molecule. This configuration is shown in Fig.~S2 in the Supplemental Material. The antibonding $\pi ^{*}$ of $\rm O_{splitt}$ lies in the band gap of \ce{ZnGa2O4} and is completely occupied in the neutral charge state. A situation similar to that is observed in \ce{In2O3} and \ce{ZnO} \cite{Janotti_165202_2007,Chatratin_074604_2019}. Due to the occupation of $\pi ^{*}$, the bond length of \ce{O2} molecule in \ce{ZnGa2O4} is larger than the isolated \ce{O2} one, and similar to the \ce{O2^{2-}} configuration. It is important to note that $\rm O_{splitt}$ is stable only in the neutral charge state and has high formation energy. Therefore, even if $\rm O_{splitt}$ is formed in \ce{ZnGa2O4}, this defect will not contribute for $n$- or $p$-type conductivity. 

The second configuration for O interstitial in the 16c site, $\rm O_{i}$, has a linear bond with two Zn atoms, forming Zn-O-Zn structure with an angle of 179$^o$. This configuration is stable in two charge states: neutral and 1-, with a transition level of 2.53~eV below the CBM. However, similar to 
$\rm O_{splitt}$ the formation energy of $\rm O_{i}$ is very high, and we do not expect a large concentration of this defect even in the O-rich condition. This is a good point for a $n$-type conductivity because $\rm O_{i}$ act as a source of electron compensation in \ce{ZnGa2O4}. 

\subsection{Source of $p$-type conductivity}

\begin{figure}
\centering
\includegraphics[width=1.0\linewidth]{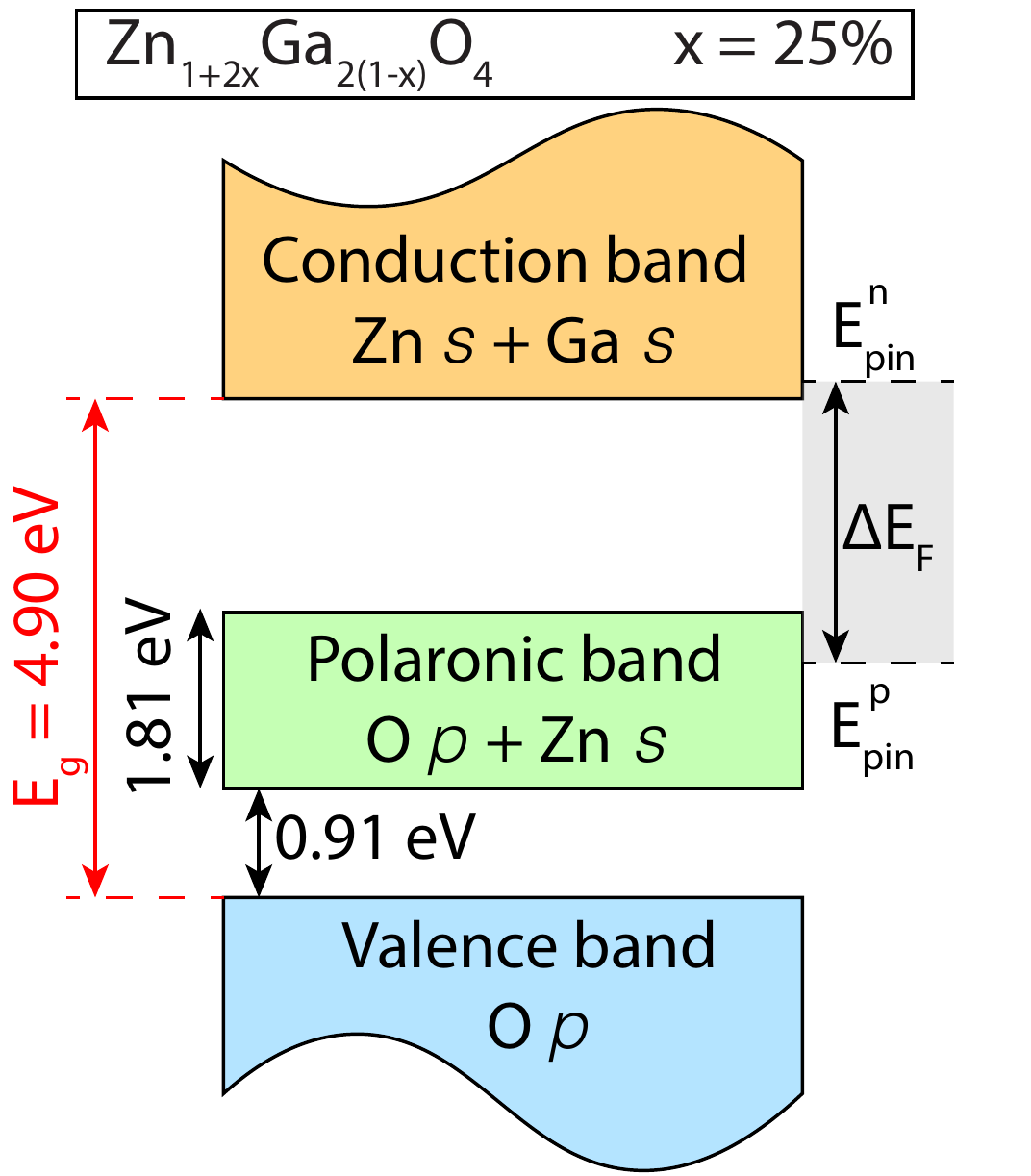}
\caption{
Schematic bands diagram for \ce{Zn_{(1+2x)}Ga_{2(1-x)}O4} alloy with $x = 0.25$. The detached band, shown in green color, is called polaronic 
band and is composed by hybridization of O 2$p$ and Zn 3$d$ orbitals. The maximum and minimum variation of Fermi level
is shown in the gray area on the right side, with the pinning for $p$-type conductivity, $E_{pin}^{p}$, crossing the polaronic band.
}
\label{polaronic_band}
\end{figure}

According to our calculations, intrinsic $p$-type conductivity is unlikely in \ce{ZnGa2O4} due to three reasons: 
(i) the holes tend to localize in the form of small polarons due to the localized character of the valence band. These polarons are very stable; 
(ii) There are no intrinsic shallow acceptors with small formation energy in the vicinity of the valence band, which creates a problem
for holes generation; and 
(iii) The Fermi level pinning for $p$-type, $E_{pin}^{p}$, lies around 2 eV above the VBM, and therefore the Fermi level cannot 
be moved towards the VB without hole compensation by intrinsic donor defects. 
Despite all these points, experimental results indicate a possible low $p$-type conductivity in high temperature, under O-rich conditions, 
with a hole density that does not exceed $\rm 10^{15}$~cm$^{-3}$ \cite{Chen_2079_2020,Chikoidze_2535_2020,Chi_2542_2021}. 

To investigate this condition, we suppose an alloy of \ce{ZnGa2O4} with an excess of Zn, forming \ce{Zn_{(1+2x)}Ga_{2(1-x)}O4}. These extra Zn atoms occupy the Ga site, forming $\rm Zn_{Ga}$.
The calculations were performed for a concentration $x = 0.25$, i.e., an increase of 25\% of the Zn atoms in the system. The excess of Zn atoms creates an intermediate band 
that is semi-occupied, assuming all $\rm Zn_{Ga}$ are in the neutral charge state. However, to determine the characteristic of the intermediate band, we suppose a completely unoccupied band with the Fermi energy at the top of host O $2$p band.
The detached band is composed by the hybridization of O 2$p$-orbitals and Zn 3$d$-orbitals (orbitals of the extra Zn in the system), forming a charge distribution similar 
to the polaron discussed previously in the point defect $\rm Zn_{Ga}$ and shown in Fig.~S3. The calculated density of states of this system is shown in the Supplemental Material Fig.~S4. Therefore, we can address the detached band to a polaronic nature. The formation of a detached band is observed in other alloys of oxides, such as \ce{In2O3} with Bi, even though in that case the nature of the detached band is not a polaronic one \cite{Sabino_034605_2019,Xufen_115205_2021}. Also, intermediate bands
play important roles in anti-doping processes in quantum materials \cite{Liu_106403_2019}, and have also been proposed for the construction of high efficiency solar cells.\cite{Luque_5014_1997,Baquio_382_2019}

\begin{figure*}
\centering
\includegraphics[width=0.80\linewidth]{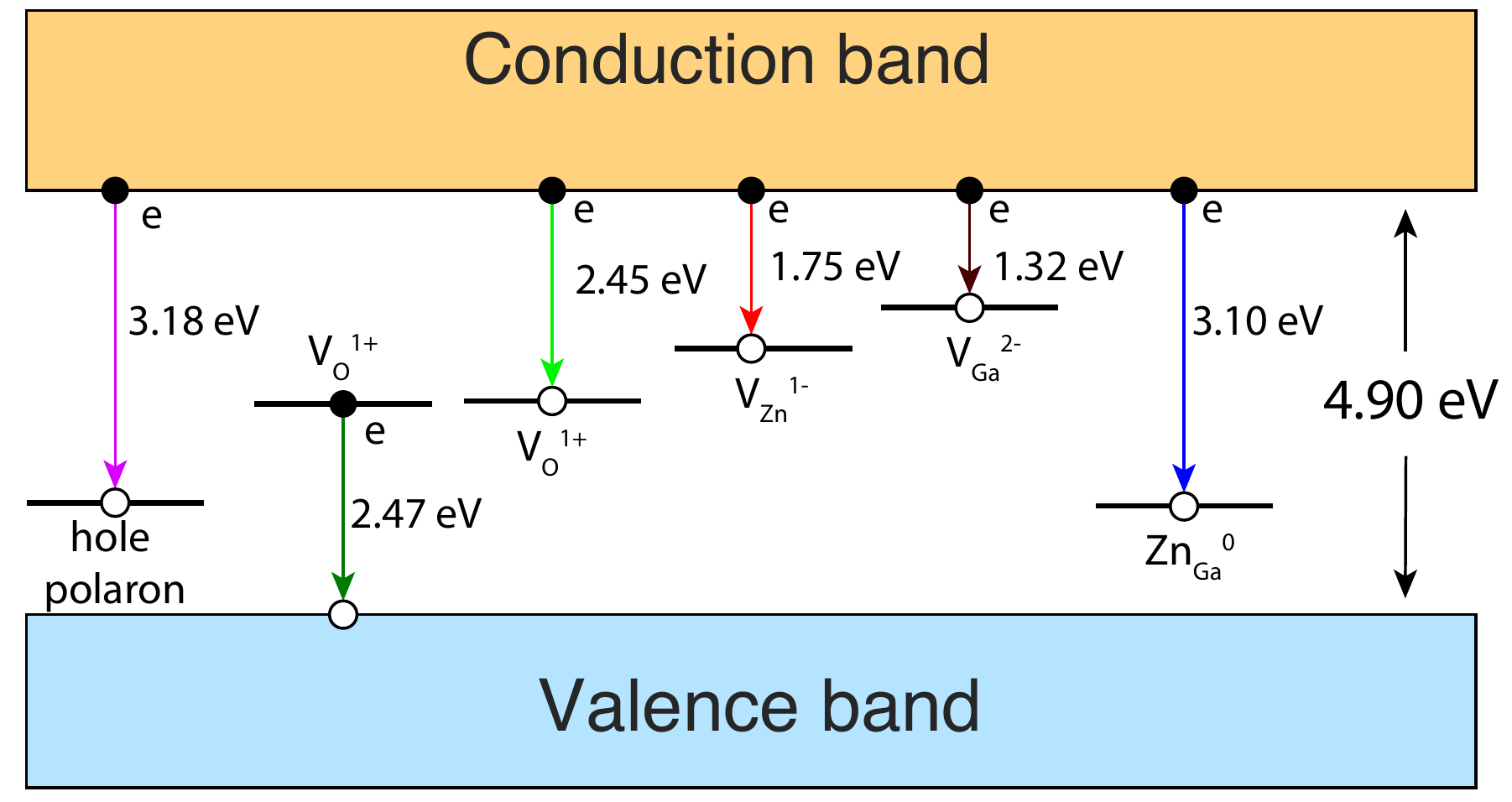}
\caption{
Optical emission of the most stables intrinsic defects of \ce{ZnGa2O4} calculated according to the Franck-Condon effect in the generalized 
coordination diagram. The filled (open) circle represents the electron (hole). 
}
\label{emission}
\end{figure*}

As we can observe in Fig.~\ref{polaronic_band}, the polaronic band lies around 0.91 eV above the host valence band maximum and has a width of 1.81 eV. The pinning of Fermi level for $p$-type conductivity, $E_{pin}^{p}$ lies inside this band. The possible Fermi level variation is shown in the gray area on the right side of Fig.~\ref{polaronic_band}. This means that is possible to shift down the Fermi level to create holes in the polaronic band before the intrinsic donor defects $\rm Ga_{Zn}$ start to form spontaneously and start to compensate the holes. In other words, it is possible to have a $p$-type conductivity inside the polaronic band. Despite that, the conductivity should occur by hoping of the localized hole in O and Zn atoms, which can lead to low charge mobility. At the same time, if the concentration of Zn in the alloy \ce{Zn_{(1+2x)}Ga_{2(1-x)}O4} decreases, the width of the polaronic band also decreases and this creates a limit for holes density in the system \cite{Sabino_034605_2019,Xufen_115205_2021}. Thus, the width and the nature of the detached band in \ce{ZnGa2O4}, grown in O-rich condition, explains the low charge mobility and the high temperature to observe this conductivity. Therefore, we can say that the $p$-type conductivity occurs in the detached polaronic band formed by an excess of Zn atoms in the system and does not occur in the valence band as suggested previously \cite{Chen_2079_2020,Chikoidze_2535_2020}.

\subsection{Optical emission}

In the optical measurement, the peaks in the photoluminescence spectrum could be a fingerprint of the defects states in the material.
Considering that, we calculate the emission photon energy for the most stable defects in \ce{ZnGa2O4}: hole polaron, $\rm V_{O}$, $\rm V_{Ga}$, $\rm V_{Zn}$, $\rm Zn_{Ga}$, $\rm Ga_{Zn}$. It is important to mention that the emission spectrum was calculated according to the Franck-Condon approximation in the generalized coordination diagram, a procedure applied before for other oxides \cite{Chatratin_074604_2019,Janotti_085202_2014}. According to the Fermi level pinning, $E_{pin}^n$ and $E_{pin}^p$, \ce{ZnGa2O4} can have the Fermi levels between the middle of the band gap and the CBM, and we considered that range to calculate the optical emission spectrum for defects. In this range, only $\rm V_{O}$ has two charge states with high stability: 2+ and 0, and therefore we considered both charge configurations. For all the remaining defects, $\rm V_{Ga}$, $\rm V_{Zn}$, $\rm Zn_{Ga}$ and $\rm Ga_{Zn}$, we considered only the most stable charge state.

If a photon with energy larger than the band gap is inserted in the semiconductor, one electron from the valence band is promoted to the conduction band leaving a hole behind. Because of the high stability of small hole polaron in \ce{ZnGa2O4}, the atoms around one O relax to localize the hole in the form of a small polaron, resulting in a quasi-particle state inside the band gap. After atomic relaxation, a free electron in the conduction band can be attracted by the small polaron and recombine, emitting light. In \ce{ZnGa2O4}, this recombination emits a photon with an energy of 3.18~eV, which corresponds to the edge of the visible spectrum, close to the ultra-violet region. This recombination process is shown in Fig.~\ref{emission}.  

For $\rm V_{O}$, two processes can occur according to the Fermi level position. First, if $E_F$ is deeper and lies in a region where $\rm V_{O}^{2+}$ is the most stable charge state, the incidence of a photon can excite one electron from the valence band to the quasi-particle level of $\rm V_{O}^{2+}$ that lies in resonance in the conduction band as discussed before. 
This procedure change the oxidation state from $\rm V_{O}^{2+}$ to $\rm V_{O}^{1+}$ 
according to the reaction: $\rm V_{O}^{2+} + 1e \rightarrow V_{O}^{1+}$. 
After the atomic relaxation, $\rm V_{O}^{1+}$ becomes unstable and tends to return to $\rm V_{O}^{2+}$ configuration. This procedure occurs by recombination of one electron of $\rm V_{O}^{1+}$ and one hole in the valence band, emitting a photon with energy equal to 2.47~eV. This energy lies in the visible spectrum and corresponds to the green color. 

The second procedure that can occur with $\rm V_{O}$ is when the $E_F$ is closer to the CBM. Under this configuration, the neutral oxygen vacancy is the most stable one, $\rm V_{O}^0$. The incidence of photons in the material can promote one electron of $\rm V_{O}^0$ to the conduction band, and promote the reaction: $\rm V_{O}^{0} + 1h \rightarrow V_{O}^{1+}$. Again, the $\rm V_{O}^{1+}$ configuration is not stable, and there is a tendency to capture one electron and return to the neutral charge state. This procedure leads to a photon emission with an energy of 2.45~eV, which is very similar to the condition before. Therefore, we expect that all the emissions associated with $\rm V_{O}$ occur in the green color independent of the Fermi level position, with both conditions shown in Fig.~\ref{emission}. The green emission is observed by experimentalists with photon energy around 2.48 eV (499 nm) \cite{Horng_6071_2017}.

Considering the other two vacancies, $\rm V_{Zn}$ and $\rm V_{Ga}$, the procedure is very similar to the second condition of $\rm V_{O}$. A photon promotes one electron that lies in the defect state of $\rm V_{Zn}^{2-}$ ($\rm V_{Ga}^{3-}$) to the conduction band, and resulting in the reaction:
$\rm V_{Zn}^{2-} (V_{Ga}^{3-}) + 1h \rightarrow V_{Zn}^{1-} (V_{Ga}^{2-})$. The excited state tends to return to the ground state and recombine with a free electron from the conduction band. This procedure emits a photon with energy equal to 1.75 eV for $\rm V_{Zn}$ and 1.32~eV for $\rm V_{Ga}$, which correspond to a spectrum with red color and at the infra-red region, respectively.

For antisite point defects, we observe different behavior. For $\rm Ga_{Zn}$, when the photon promotes one electron from the valence band to the $\rm Ga_{Zn}^{1+}$ defect level, the system relaxes to $\rm Ga_{Zn}^{0}$ according to the reaction: $\rm Ga_{Zn}^{1+} + 1e \rightarrow Ga_{Zn}^{0}$. The defect level in the neutral condition lies in resonance in the conduction band, and therefore this condition favors recombination through band-band emission. In other words, $\rm Ga_{Zn}$ will not have a photoemission fingerprint because the band-band is the lowest energetic path. 

For $\rm Zn_{Ga}$, the incident light in \ce{ZnGa2O4} promotes one electron from $\rm Zn_{Ga}^{1-}$ state to the conduction band by the process: $\rm Zn_{Ga}^{1-} + 1h \rightarrow Zn_{Ga}^{0}$. If the Fermi level is close to the CBM, $\rm Zn_{Ga}^{0}$ is unstable and tends to capture one free electron to return to the original charge state. A photon with energy equal to 3.10~eV, which is close to the UV spectrum, is emitted in this procedure. According to the emission spectrum fingerprint, we can say that hard to distinguish 
the hole polaron and $\rm Zn_{Ga}$ defect.

\section{Summary and Conclusions}

In summary, we calculate the electronic, optical, and defect properties in \ce{ZnGa2O4}, a potential candidate for high-energy power electronics and UV-bling detectors. 
We observe that antisites play important role in the conductivity of \ce{ZnGa2O4}. While $\rm Ga_{Zn}$ is likely the main source of unintentional $n$-type conductivity, $\rm Zn_{Ga}$ is the most stable acceptor defect and the main source of electron compensation. The $p$-type conductivity in the valence band is difficult to be achieved. There is a large tendency for hole localization in the form of small polarons, either as self-trapped holes or bound to intrinsic defects, pinning the Fermi level far above the VBM.

We conclude that $p$-type conductivity in \ce{ZnGa2O4} could be achieved only with excess Zn atoms. This will form an intermediate polaronic band inside the band gap of the material. In this band, the 
conductivity occurs by hopping, which explains the low mobility and low hole density observed in experimental results. In addition, we compute 
the emission spectrum of the lowest energy defects in \ce{ZnGa2O4}. Most of the defects that are associated with polarons emit in the spectrum close to ultra-violet, while the oxygen vacancy emits in the green. The possibility for both types of conductivity, better thermal dissipation, and isotropic properties make \ce{ZnGa2O4} a potential candidate to replace \ce{Ga2O3}, or at least complements it as an ultra-wide band gap material for electronic and optoelectronic applications.

\section{Acknowledgment}

FPS and GMD thank FAPESP (grants 2019/21656-8 and 17/02317-2), CNPq and CAPES for financial support, and the National Laboratory for Scientific Computing (LNCC/MCTI, Brazil) for providing HPC resources on the SDumont supercomputer. Work of AJ and IC was supported by the NSF Early Career Award grant no. DMR-1652994, the Extreme Science and Engineering Discovery Environment (XSEDE) supported by National Science Foundation grant number ACI-1053575, and the Information Technologies (IT) resources at the University of Delaware.

%\bibliography{/Users/fernando/Dropbox/pos_doc/jshort_25Feb2015.bib,/Users/fernando/Dropbox/pos_doc/boxref_25Feb2015.bib}

%\bibliography{Bib/jshort_25Feb2015.bib,Bib/boxref_25Feb2015.bib}
%\bibliography{jshort_25Feb2015.bib,boxref_25Feb2015.bib}

%\bibliography{jshort_25Feb2015.bib,boxref_25Feb2015.bib}


\end{document}